\documentclass[preprint,showpacs,showkeys]{revtex4-1}
\usepackage{amsmath}
\usepackage{graphics}
\usepackage{graphicx}
\newcommand{\ts}{\theta_S}
\begin{document}

% Use the \preprint command to place your local institutional report
% number in the upper righthand corner of the title page in preprint mode.
% Multiple \preprint commands are allowed.
% Use the 'preprintnumbers' class option to override journal defaults
% to display numbers if necessary
%\preprint{}

%Title of paper
\title{Type 1 2HDM as  effective theory of supersymmetry}

% repeat the \author .. \affiliation  etc. as needed
% \email, \thanks, \homepage, \altaffiliation all apply to the current
% author. Explanatory text should go in the []'s, actual e-mail
% address or url should go in the {}'s for \email and \homepage.
% Please use the appropriate macro foreach each type of information

% \affiliation command applies to all authors since the last
% \affiliation command. The \affiliation command should follow the
% other information
% \affiliation can be followed by \email, \homepage, \thanks as well.
\author{Hua.~Shao}
\email[]{shaohua@itp.ac.cn}
%\homepage[]{Your web page}
%\thanks{}
%\altaffiliation{}
\affiliation{State Key Laboratory of Theoretical Physics,
Institute of Theoretical Physics, Chinese Academy of Sciences,
P.O. Box 2735, Beijing 100190, P.R. China}

%Collaboration name if desired (requires use of superscriptaddress
%option in \documentclass). \noaffiliation is required (may also be
%used with the \author command).
%\collaboration can be followed by \email, \homepage, \thanks as well.
%\collaboration{}
%\noaffiliation

\date{\today}

\begin{abstract}
It is generally believed that the low energy effective theory of the minimal supersymmetric standard model is the type 2 two Higgs doublet model.
We will show that the type 1 two Higgs doublet model can also as the effective of supersymmetry in a specific case with high scale supersymmetry breaking and gauge mediation.
If the other electroweak doublet obtain the vacuum expectation value after the electroweak symmetry breaking, the Higgs spectrum is quite different.
A remarkable feature is that the physical Higgs boson mass can $125~\textmd{GeV}$ unlike in the ordinary models with high scale supersymmetry in which the Higgs mass is generally around $140~\textmd{GeV}$.
\end{abstract}

% insert suggested PACS numbers in braces on next line
\pacs{12.60.Jv}
% insert suggested keywords - APS authors don't need to do this
\keywords{Higgs mass; supersymmetry; gauge mediation}

%\maketitle must follow title, authors, abstract, \pacs, and \keywords
\maketitle

% body of paper here - Use proper section commands
% References should be done using the \cite, \ref, and \label commands

\section{Introduction}
The two Higgs doublet model~(2HDM)\cite{Lee:1973iz} is one of the minimal extensions of the Standard Model.
It has many remarkable features, especially on discussions to the CP-violation\cite{Lee:1974jb,Liu:1987ng,Wu:1994ja} and baryon asymmetry in physics\cite{Cline:1996mga,Cline:1995dg}.
There are too many free parameters in the Higgs sector of the 2HDM, it is important to find an underlying theory of it to fix some these free parameters.

The Minimal supersymmetric Standard Model~(MSSM)\cite{Nilles:1983ge,Haber:1984rc} is another one most favored candidates of the Standard Model.
If all of the supersymmetric partners of the known particles are taken to be very heavy, the Higgs sector of the MSSM takes the Type 2 2HDM as its low energy effective theory in which one Higgs doublet couples to up type quarks and the other one couples to down type quarks and leptons.
In this case the quartic couplings of Higgs doublets are highly restricted by supersymmetry.
It is well known that the lightest Higgs boson is lighter than the $Z$ gauge boson at tree level.
To lift the mass of the lightest Higgs mass, large quantum corrections are needed\cite{Ellis:1990nz,Okada:1990vk,Haber:1990aw}.

It is possible that supersymmetry breaks at high energy and the weak scale is obtained by fine-tuning.
Many excellent works have been proposed under the assumption of high scale supersymmetry breaking\cite{ArkaniHamed:2004yi,Giudice:2004tc,Liu:2005rs}.
There is only one light Higgs doublet  at the weak scale and detailed studies show that its natural mass will be around $140~\textmd{GeV}$\cite{Hall:2009nd}.
If we consider the exceed of events at $126~\textmd{GeV}$ in ATLAS\cite{ATLAS} and $124~\textmd{GeV}$ in CMS\cite{CMS} as the signals of the Higgs boson, theories with high scale supersymmetry breaking seem to be wrong due to the largeness of quantum corrections from the top quark.

In this work we will show that in a special supersymmetric case, there will be two electroweak bosonic doublets at the weak scale.
One of them comes from fine-tuning as in the high scale supersymmetry breaking and couple to all of the fermions to generate their masses after the electroweak symmetry breaking and the other one comes from high scale dynamics which does not couples to Standard Model fermions.
They interact with each other and both of acquire vacuum expectation values~(VEV) to contribute the electroweak symmetry breaking.
The effective theory at the weak scale is the type 1 2HDM in which one of the quartic coupling constant can be predicted by supersymmetry.
The lightest Higgs boson can be as low as $125~\textmd{GeV}$ unlike the ordinary models with high scale supersymmetry breaking since there are light particles except the Standard Model Higgs field.

This paper is organized as followings.
In the next section the basic structure of the model is established, and discuss the  possible underlying dynamics.
In section 3, we will focus on the  properties of the Higgs boson, especially on its mass.
A brief summary will be given in the last section.

\section{Type 1 2HDM and supersymmetry}
It is generally believed that the Higgs sector of the supersymmetric Standard Model  is the type 2 2HDM for weak scale supersymmetry  or the one Higgs as in the Standard Model  for fine-tuned high scale supersymmetry.
In this section, it will be shown that under some special assumptions in theories gauge mediated supersymmetry breaking, one of the doublets of the bosonic components of messengers   is accidently as light as the weak scale.
The effective theory will be the type 1 2HDM.

In theories with high scale supersymmetry breaking, to generate the electroweak scale one has to fine-tune the Higgs mass matrix
\begin{equation}\label{mass matrix in mssm}
M_H^2=\left(
        \begin{array}{cc}
          m_{H_u}^2+|\mu|^2 & B\mu \\
          B\mu & m_{H_d}^2+|\mu|^2 \\
        \end{array}
      \right)
\end{equation}
to have the negative mass square eigenvalue of the order of the weak scale. The physical Higgs mass is determined by the quartic coupling constant of the linear combination of the light eigenstate. The dominant contributions to its mass is from gauge couplings and top quark Yukawa interactions.

In a complete theory the mechanisms that break supersymmetry  and mediate the supersymmetry breaking effects to the visible sector are needed.
Theories with gauge mediated supersymmetry breaking~(GMSB)\cite{Dine:1993yw,Dine:1994vc,Giudice:1998bp,Meade:2008wd} are well studied in the past.
The supersymmetry breaking effects are transmitted into the visible sector through the Standard Model gauge interactions.
It is known that if there are only gauge interactions between messengers and MSSM particles, the bilinear mixing breaking term $B\mu$ cannot be generated up to two loops level.
The bilinear breaking term $B\mu$ plays essential role in the electroweak symmetry breaking so it is required to be generated in some other ways.
In another words in additional to gauge interactions, there must be other direct couplings between the Higgs doublets in the MSSM and messengers.
Ref.\cite{Komargodski:2008ax} studied various couplings and their effects in the most general ways.

In the past the fields which  break  supersymmetry and mediate the effects are usually taken to be hidden.
They are generally heavy and interact with visible particles indirectly.
Here we consider a different case in which some of the messengers are light enough to be observed.
The simplest way is to assume the supersymmetry breaking effects to some messenger fields are too strong to make light particles.
For instance  consider a pair of messenger $\phi$ and $\overline{\phi}$, they interact with background supersymmetry breaking field as
\begin{equation}\label{gauge mediation}
W=X\Phi\overline{\Phi},\qquad \langle X\rangle=M+\theta^2F.
\end{equation}
The fermionic components of messengers form a Dirac fermion with mass $M$ and the degeneracy between bosons and fermions are violated  by the supersymmetry breaking effect $F$. The mass squares  of two bosonic messengers are
\begin{equation}\label{mass relation}
m_{\pm}^2=M^2\pm F.
\end{equation}
The gauge mediated soft masses are of the order of
\begin{equation}
M_S\sim\frac{g^2}{16\pi^2}\frac{F}{M},
\end{equation}
where $g$'s are the gauge coupling constants.
In the extreme case that $F\sim M^2$ one of the bosonic messengers can be very light  and comparable  with the weak scale.

Suppose that one of the electroweak doublet messengers which are the eigenstate of the smaller eigenvalue of eq.~(\ref{mass relation}) are light   and couple to the Higgs doublets to generate $B\mu$, we will have a light doublet at the weak scale and interact with the Higgs doublet.
In addition to the fine-tuned light Higgs doublet $h$, there is also a doublet $\phi$ with the same quantum as $h$ and its mass
\begin{equation}\label{lighter messenger mass}
m_\phi^2=M^2-F\sim m_{EW}^2.
\end{equation}
The low energy effective theory is just the type 1 2HDM.
The light fields are
\begin{equation}
h=H_u\sin\beta+\widetilde{H}_d^*\cos\beta
\end{equation}
and
\begin{equation}\label{definition of phi}
\phi=\frac{1}{\sqrt{2}}(\Phi+\overline{\Phi}^*),
\end{equation}
where   $\Phi$ and $\overline{\Phi}$ in eq.~\ref{definition of phi} indicate the  bosonic components of superfields  $\Phi$ and $\overline{\Phi}$ in eq.~(\ref{gauge mediation}), respectively.

The interactions between $h$ and $\phi$ are model dependent, they are determined by the underlying theory which  generate all the breaking terms.
For practice they are parameterized as free parameters.
The scalar potential is chosen as
\begin{equation}\label{scalar potential}
V=m_h^2|h^2|+m_{h\phi}^2h\phi^+m_{\phi}^2|\phi^2|+\lambda_h|h|^4+\lambda_{h\phi}|h|^2|\phi|^2+\lambda_{\phi}|\phi|^4+c.c.
\end{equation}
There are three comments to the potential:
\begin{itemize}
  \item In the potential~(\ref{scalar potential}) all  of the dimensional parameters $m_{h}^2$, $m_{\phi}^2$ and  $m_{h\phi}^2$ are  assumed at the weak scale although their origins are quite different.
    $m_{h}^2$ is the smaller eigenvalue of the mass square matrix~(\ref{mass matrix in mssm}) which is obtained by the fine-tuning and  $m_{\phi}^2$ is the lighter bosonic messenger mass~(\ref{lighter messenger mass}).
    Their mixing mass $m_{h\phi}^2$ is the net effect in determining $\mu$ and $B\mu$ which is need to be explained in details.
    Suppose that there are couplings between the Higgs fields and the hidden sector doublets as part of the underlying mechanism \footnote{The couplings between Higgs and messenger fields are not clear for us so all possible couplings are assumed to  exist. }to  generate $\mu$ and $B\mu$
      \begin{equation}
      W=\alpha_iH_uO_d^i+\beta_iH_dO_u^i+W(O_{u,d}),
      \end{equation}
      in addition to bilinear mixing $B\mu$, bilinear soft breaking couplings between Higgs fields and messenger fields are  often generated by the same interactions.
      If one bosonic components of the messenger field is accidentally light there is bilinear mixing mass $m_{h\phi}^2$ in general.
      It will be shown in the next section the mixing mass term plays crucial role in this model.
  \item The quartic coupling $\lambda_h$ in eq.~(\ref{scalar potential}) can be predicted in terms of gauge and Yukawa interactions as in the ordinary high scale supersymmetric theories.
      \begin{equation}
      \lambda_h=\frac{g^2+g'^{2}}{8}\cos^22\beta+(\textrm{quantum corrections}).
      \end{equation}
      Notice that for non-zero mixing mass $m_{h\phi}^2$ it is generally that the VEV of $\phi$ is non vanishing. The Yukawa couplings here are different from those in the Standard Model. Unlike $\lambda_h$, the quartic  coupling $\lambda_{\phi}$ in eq.~(\ref{scalar potential}) is not necessary related with the gauge coupling constants. To see this we can calculate the D-term  of $\phi$ and its orthogonal field
      \begin{equation}
      \phi_H=\frac{1}{\sqrt{2}}(\Phi-\overline{\Phi}^*),
      \end{equation}
      it is easily obtained that
      \begin{equation}
      D\supset |\Phi|^2-|\overline{\Phi}|^2=\phi\phi_H^++c.c.
      \end{equation}
      It can be seen that in the D-term $\phi$ always couples to its heavy orthogonal fields $\phi_H$.
      It is believed that such an interaction is irrelevant at low energy, it decouples from the low energy effective theory.
      The fact is analogous to the case in the MSSM with $\tan\beta=1$, the tree level quartic coupling of the lightest Higgs vanishes.  The quartic couplings in eq.~(\ref{scalar potential}) can be generated if the Higgs doublets couple to a hidden sector singlet. The simplest example is
      \begin{equation}\label{singlet coupling}
      W=\alpha NH_uH_d+\beta N\phi\overline{\phi},
      \end{equation}
      the F-term potential of the singlet $N$ contain all types of quartic couplings.
      It is worth to notice that $\lambda_h$ and $\lambda_{\phi}$ are strictly positive but $\lambda_{h\phi}$ not.
      It can be negative.
      For example  $\lambda_{h\phi}<0$ when  $\alpha$ and $\beta$ in eq.~(\ref{singlet coupling}) have opposite signs.
  \item It is obviously that the potential~(\ref{scalar potential}) is not the most general invariant potential contains $h$ and $\phi$, terms like $(h^+\phi+c.c.)|h|^2$ which is also allowed by the gauge symmetries are not included. They are omitted here since they cannot be generated by simple interactions.
  \end{itemize}

If the vacuum state of the potential~(\ref{scalar potential}) is at the point with non-zero VEV of $\phi$, the Higgs sector is quite different from those in the high scale supersymmetry breaking theories.
The details of the Higgs sector will be studied in the next section.

\section{The Higgs sector, Higgs mass}
From the potential~(\ref{scalar potential}) the vacuum conditions are
\begin{equation}
\frac{\partial V}{\partial h^+}=m_h^2h+m_{h\phi}^2\phi+2\lambda_{h}|h|^2h+2\lambda_{h\phi}|\phi|^2h=0
\end{equation}
\begin{equation}
\frac{\partial V}{\partial \phi^+}=m_{\phi}^2\phi+m_{h\phi}^2h+2\lambda_{\phi}|\phi|^2\phi+2\lambda_{h\phi}|h|^2\phi=0
\end{equation}
The electroweak symmetry will break  for suitable choice of parameters.

Define the VEV's for $h$ and $\phi$ as $v_h$ and $v_\emph{}{\phi}$, respectively
\begin{equation}
v_h=v\sin\theta_S,\qquad v_{\phi}=v\cos\theta_S
\end{equation}
in which $M_Z^2=\frac{g^2+g^{'2}}{2}v^2$.
The largeness of the top quark mass requires that
\begin{displaymath}\tan\theta_S=v_h/v_{\phi}\end{displaymath}
can not be too small.

The Higgs sector in the 2HDM has been widely discussed, here we will focus on the lightest CP-even Higgs boson mass in the decoupling limit. In the absence of CP-violation, the mass of the CP-odd Higgs boson is
\begin{equation}
M_A^2=-\frac{m_{h\phi}^2}{\sin\theta_S\cos\theta_S}.
\end{equation}
The mass matrix of the CP-even neutral Higgs bosons is
\begin{equation}
M^2=\left(
      \begin{array}{cc}
        -m_A^2\cos^2\theta_S+4\lambda_hv^2\sin^2\theta_S & (m_A^2+4\lambda_{h\phi}v^2)\sin\theta_S\cos\theta_S \\
        (m_A^2+4\lambda_{h\phi}v^2)\sin\theta_S\cos\theta_S & -m_A^2\sin^2\theta_S+4\lambda_{\phi}v^2\cos^2\theta_S \\
      \end{array}
    \right).
\end{equation}
The physical mass of the  lightest Higgs boson mass in the decoupling limit is
\begin{equation}\label{upper bound}
M_h^2=4\lambda_hv^2\sin^4\theta_S+8\lambda_{h\phi}v^2\sin^2\theta_S\cos^2\theta_S+4\lambda_{\phi}v^2\cos^4\theta_S.
\end{equation}

\begin{figure}
\centering
\includegraphics[width=0.6\textwidth]{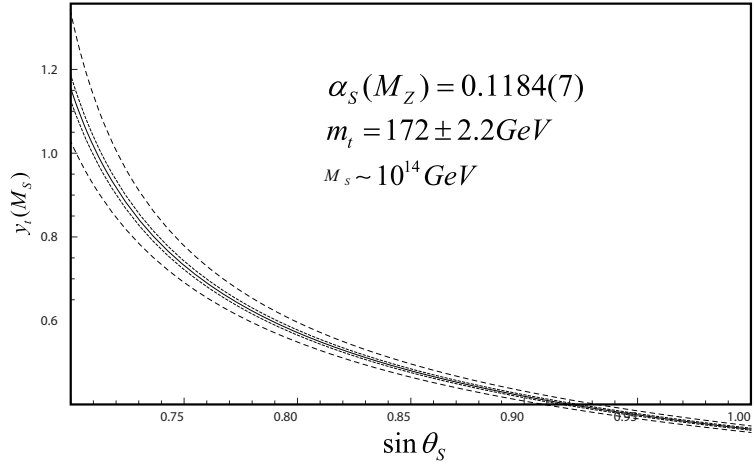}\\
\caption{The top quark Yukawa coupling at the supersymmetry breaking scale $M_S$ as a function of $\sin\ts$. }\label{ytl}
\end{figure}

As stated in the previous section the quartic coupling constant $\lambda_h$ can be predicted in terms of the Standard Model couplings. The top quark Yukawa coupling constant at the weak scale is
\begin{equation}\label{top Yukawa}
y_t(M_Z)=\frac{m_t}{v\sin\theta_S}.
\end{equation}
The one loop $\beta$-function for $y_t$ is
\begin{equation}
\frac{dy_t}{dt}=\frac{y_t}{16\pi^2}\big(\frac{9}{2}y_t^2-8g_3^2\big)
\end{equation}
The running of the $y_t$ is given in  fig.\ref{ytl}.
It can be seen that $y_t$  remains  perturbative at high scale for $\sin\theta_S>0.7$.

%%%%%%%%%%%%%%%%%%%%%%%%%%%%%
%%%%%%%%%%%%%%

\begin{figure}
\centering
\includegraphics[width=0.6\textwidth]{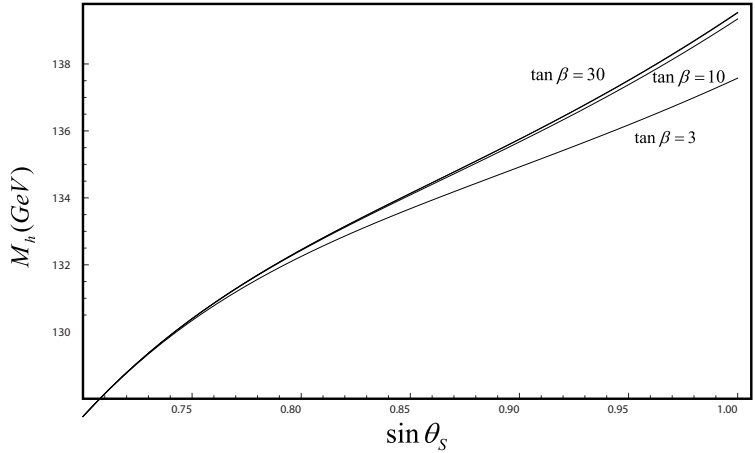}\\
\caption{the Higgs mass in the absent of contributions from $\lambda_{\phi}$ and $\lambda_{h\phi}$}\label{figure:mh}
\end{figure}

The quartic coupling $\lambda_h$ at the supersymmetry breaking scale $M_S$  is assumed to be the same as the MSSM prediction
\begin{equation}
\lambda_h(M_S)=\frac{g^2+g^{'2}}{8}(M_S)\cos^22\beta.
\end{equation}
The one loop running of $\lambda_h$ to the first order is
\begin{equation}
\frac{d\lambda_h}{dt}=\frac{3}{4\pi^2}\big(2\lambda_h^2+\lambda_h y_t^2-\frac{1}{2}y_t^4\big)
\end{equation}
The other two quartic coupling constants in eq.~(\ref{scalar potential}) are model dependent.
To the first order of approximation their contributions to the Higgs mass are neglected, the  Higgs mass~(\ref{upper bound}) becomes
\begin{equation}
m_h^2=4\lambda_hv^2\sin^4\theta_S.
\end{equation}
It seems that the Higgs mass square proportional to $\sin^4\ts$ and we will be free to reduce the Higgs mass to the observed one.
In fact the effect of the top Yukawa interaction is enhanced since eq.~(\ref{top Yukawa}).
As a function of $\sin\ts$ and $\tan\beta$ the physical Higgs mass are shown in fig.\ref{figure:mh}.
%%%%%%%%%%%%%%%%%%%%%%%%%%%%%
%%%%%%%%%%%%%%

\begin{figure}
\centering
\includegraphics[width=0.6\textwidth]{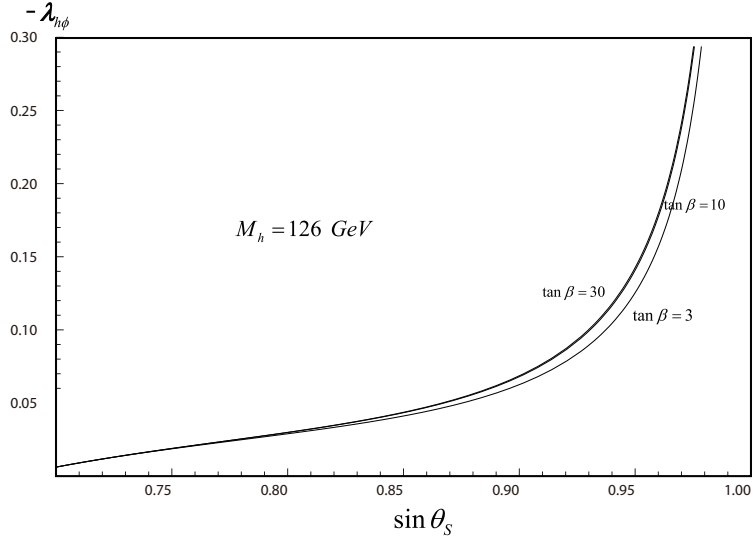}\\
\caption{ For different $\sin\ts$ and $\tan\beta$ the $\lambda_{h\phi}$ shown in the figure can make the Higgs boson to be $125~\textmd{GeV}$}\label{figure:lhp}
\end{figure}

We can see from fig.\ref{figure:mh} that the Higgs mass is smaller than the generally predicted $140~\textmd{GeV}$ in  models with high scale supersymmetry breaking.
But is still too heavy compare to the measured Higgs mass even for the lowest allowed value of $\sin\ts$.
To further reduce the Higgs mass, contributions from $\phi$ have to be considered.
As stated in the previous section, the coupling constant of the mixed quartic term $\lambda_{h\phi}$ can be negative.
It will be another source to lower the Higgs mass.
To obtain  the observed Higgs mass, the required $\lambda_{h\phi}$ is shown in fig.\ref{figure:lhp}.
Here the terms in eq.~(\ref{upper bound}) which proportional to $\cos^4\ts$ are neglected due to the smallness of it.
The requirement that the potential~(\ref{scalar potential}) bounds below is
\begin{equation}
\lambda_{h\phi}^2<4\lambda_h\lambda_{\phi}.
\end{equation}
It is easily satisfied in the presence of $\lambda_{\phi}$.

%%%%%%%%%%%%%%%%%%%%%%%%%%%%%
%%%%%%%%%%%%%%

It can been seen in the previous discussions that the Standard Model like Higgs boson is able to be at $126~\textmd{GeV}$ in a large portion of the parameter space.
In the decoupling limit $M_A^2\gg M_Z$, the properties of the lightest Higgs boson is exactly the same as the Standard Model Higgs boson.
For smaller $M_A^2$ there will be rich phenomena at the LHC.

\section{Conclusion}
It has been shown in this paper that under very special assumptions a supersymmetric theory takes the type 1 2HDM as its low energy effective theory.

The basic structure and possible simple microscopic theories are discussed in section 2.
In this work we choose the simplest potential to the further studying, in fact the underlying theory can be more complicate  to afford other terms that do not included in eq.~(\ref{scalar potential}).
It is possible that these terms can make the VEV of $\phi$ has an irreducible phase with respect to $h$ to introduce a new source of CP-violation to the theory.  These effects will be discussed in our future studies.

The phenomena of the Higgs bosons is the same as the type 1 2HDM.
These objects have been well studied\cite{Branco:2011iw}.
If the type 1 2HDM is considered as effective theory of supersymmetry, one quartic coupling can be predicted unlike the usual 2HDM in which all couplings are free.
As shown in the previous section, the mass of the Standard Model like Higgs boson is easy to be at $125~\textmd{GeV}$.

I would like to thank  Minkai.~Du, Chun.~Liu, Jiashu.~Lu, Shuo.~Yang and Zhenhua.~Zhao for very fruitful  discussions.
This work
was supported in part by the National Natural Science Foundation of
China under nos. 11075193 and 10821504, and by the National Basic
Research Program of China under Grant No. 2010CB833000.

\end{document}